\pdfoutput=1
\documentclass[%
 letterpaper,twocolumn,english,
 superscriptaddress,
showpacs,preprintnumbers,
 amsmath,amssymb,
 aps,
 prc,
floatfix,
]{revtex4-1}

\usepackage[usenames]{color}
\usepackage{amsmath}
\usepackage{amssymb}
\usepackage{longtable}  %longtable
\usepackage{graphicx}% Include figure files
\usepackage{lineno,xcolor}
\usepackage[unicode=true,bookmarks=false,breaklinks,pdfborder={0
    0 0},backref=false,colorlinks=true]{hyperref}  
\hypersetup{pdftitle={Parity Violation in Deep Inelastic Scattering with the SoLID Spectrometer 
	at JLab},pdfauthor={Y. X. Zhao , et. al. (The SoLID Collaboration)},pdfsubject={v0.0: nucl-ex,
    hep-ex},pdfkeywords={PVDIS, DIS, weak mixing angle, weak couplings},linkcolor=DarkBlueCite, citecolor=DarkBlueCite,
  urlcolor=DarkBlueCite}  
\usepackage[hyphenbreaks]{breakurl}

\makeatletter

\usepackage{dcolumn}% Align table columns on decimal point
\usepackage{bm}% bold math

\definecolor{DarkBlueCite}{rgb}{0.1,0.0,0.5}

\makeatother

\begin{document}


\author{Y. X.~Zhao}\email[Corresponding author: ]{yuxiang.zhao@stonybrook.edu}
\affiliation{Stony Brook University, NY, 11794, USA \\
For the SoLID Collaboration}
\collaboration{The SoLID Collaboration}
\noaffiliation

\preprint{APS/123-QED}
\title{ Parity Violation in Deep Inelastic Scattering with the SoLID Spectrometer at JLab} 

\date{\today}
%\vspace*{0.1cm}

\begin{abstract}
Measurements of parity-violating asymmetries in DIS region using 
the SoLID spectrometer at Jefferson Lab (JLab) Hall A in the 12 GeV era are presented. A proposal with a polarized electron 
beam on unpolarized deuteron and proton targets has been approved with an A rating by the JLab PAC. The 
deuteron measurement aims to measure the weak mixing angle $\sin^2 \theta_W $ with a precision 
of $\pm$ 0.0006 as well as to access the fundamental coupling constants $C_{2q}$ with a high 
precision. This measurement is ideally suited for testing the Standard Model with the potential to 
probe charge symmetry violation and resolve the quark-quark correlations in the DIS 
region. The proton experiment provides a clean measurement of $d/u$ ratio in the high-$x$ 
region free of nuclear corrections. To achieve these goals, the SoLID spectrometer was proposed and designed to handle a 
high luminosity with a large acceptance.  In 
this article, the details of the approved measurements are discussed, along with 
new ideas with PVDIS using a polarized $^3$He target to access new $\gamma-Z$ interference 
polarized structure functions and a unpolarized $^{48}$Ca target to study the EMC effect.  
\end{abstract}


\maketitle
%\vspace{5.0cm}

%xxx.xx Introduction. xxx.xx
%__________________________introduction part______________________________
\section{Introduction}
Symmetries play a central role in physics. Parity, time reversal, and charge conjugation symmetries etc. were
naturally assumed to be conserved until T.D. Lee and C.N. Yang first suggested parity violation \cite{PhysRev.104.254}. 
C.S. Wu led the first experiment in nuclear $\beta$ decay which confirmed the parity violation \cite{PhysRev.105.1413}.
The Nobel Prize in physics was awarded to Lee and Yang in 1957 ``for their penetrating investigation 
of the so-called parity laws which has led to important discoveries regarding the elementary particles."
In the following decade after the first observation of parity violation, many models/theories were invented
to explain the phenomenon. Among them is the Glashow-Weinberg-Salam (GWS) theory \cite{Glashow:1961tr, Weinberg:1967tq, Salam:1968rm} 
which yields the unification of electroweak interaction and predicts a new electrically-neutral boson $Z^{0}$. In GWS theory, all spin-1/2
particles carry two types of couplings: axial and vector couplings. The axial coupling $g_A$ describes the
difference of the strength of neutral-weak interaction for the left- and right-handed states of spin-1/2 particles,
while the vector coupling $g_V$ describes the average of the two. For pure virtual photon exchange, there is no
difference for left- and right-handed particles, hence only vector coupling exists and it is equal to the electrical
charge of the particle. For the $W$ boson, it only interacts with left-handed fermions. For the $Z$ boson,
it interacts with both left- and right-handed fermions, for instance, the electron axial coupling $g_A^{e} = - \frac{1}{2}$ and
the vector coupling $g_V^{e} = -\frac{1}{2} + 2 \sin^2 \theta_W $. 

\begin{figure}%[ht]
\includegraphics[width=90mm]{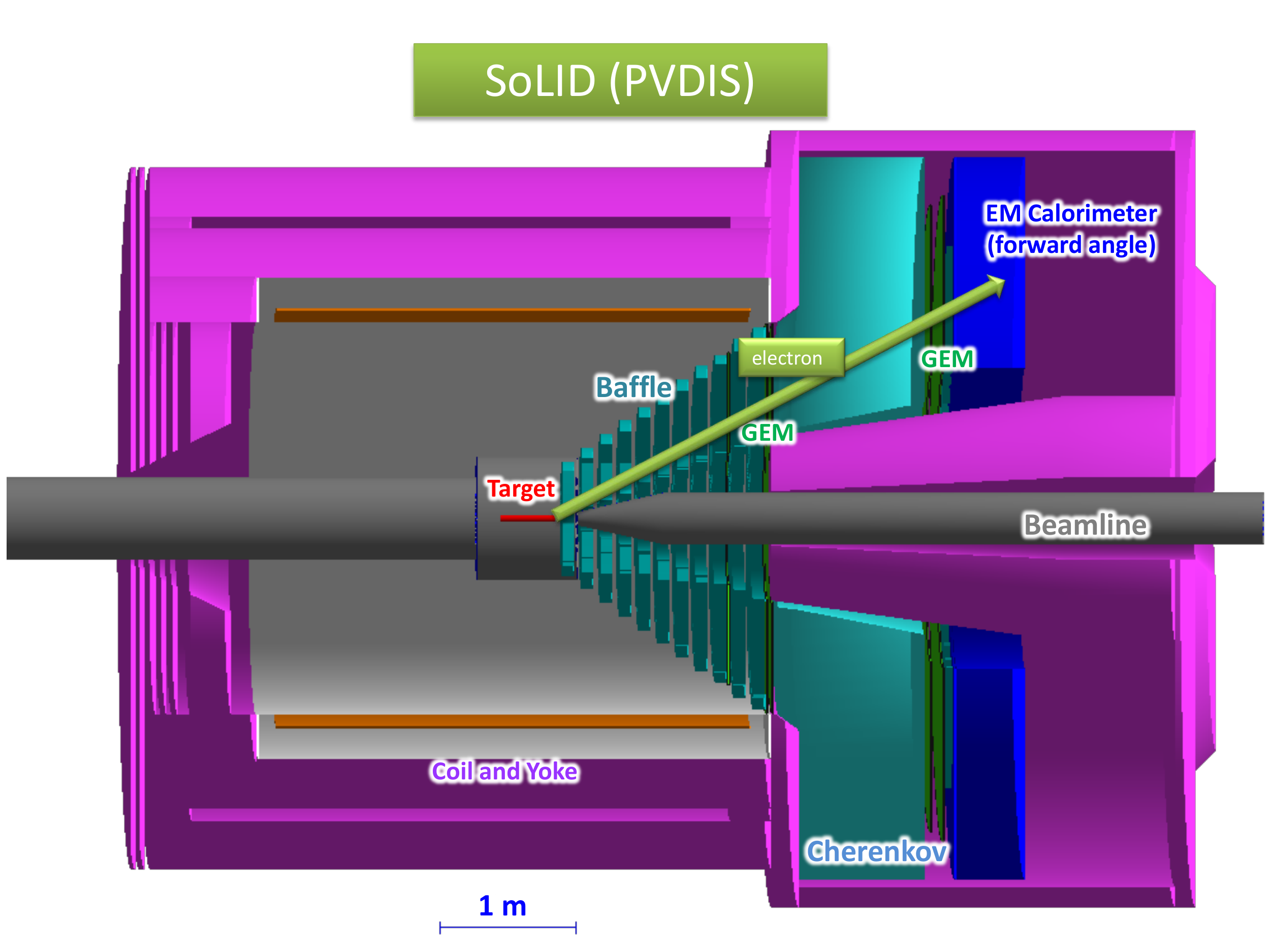}
\caption{(Color online) SoLID spectrometer for the PVDIS program.}
\label{fig:solid_pvdis}
\end{figure}

In electron-nucleon (nuclei) scattering, parity violation is usually observed by measuring the non-zero asymmetry
\begin{equation}
A_{PV}=\frac{\sigma_R - \sigma_L}{\sigma_R + \sigma_L}
\end{equation}
with longitudinally polarized electrons and unpolarized nucleon or nuclear targets. 
The beauty of the measurement, especially in the DIS region, is that the interaction vertex provides the unique 
information on effective electron-quark couplings while the quarks probed by the neutral current reveal the 
internal structure of the nucleon. It has been a powerful tool since the 1970s to access the
fundamental quantities of QCD, to study the nucleon structure, and to search for new physics beyond the Standard Model (SM).
In the JLab 12 GeV era a Solenoidal Large Intensity Device (SoLID), shown in Figure \ref{fig:solid_pvdis},
is proposed to measure the parity violating
asymmetry in the DIS region (PVDIS) using different unpolarized/polarized targets \cite{Chen:2014psa}.
Details of these measurements will be discussed in the following sections.

\section{Physics programs}
\subsection{PVDIS with longitudinally polarized electrons and a unpolarized deuteron target}
\begin{figure}%[ht]
\includegraphics[width=90mm]{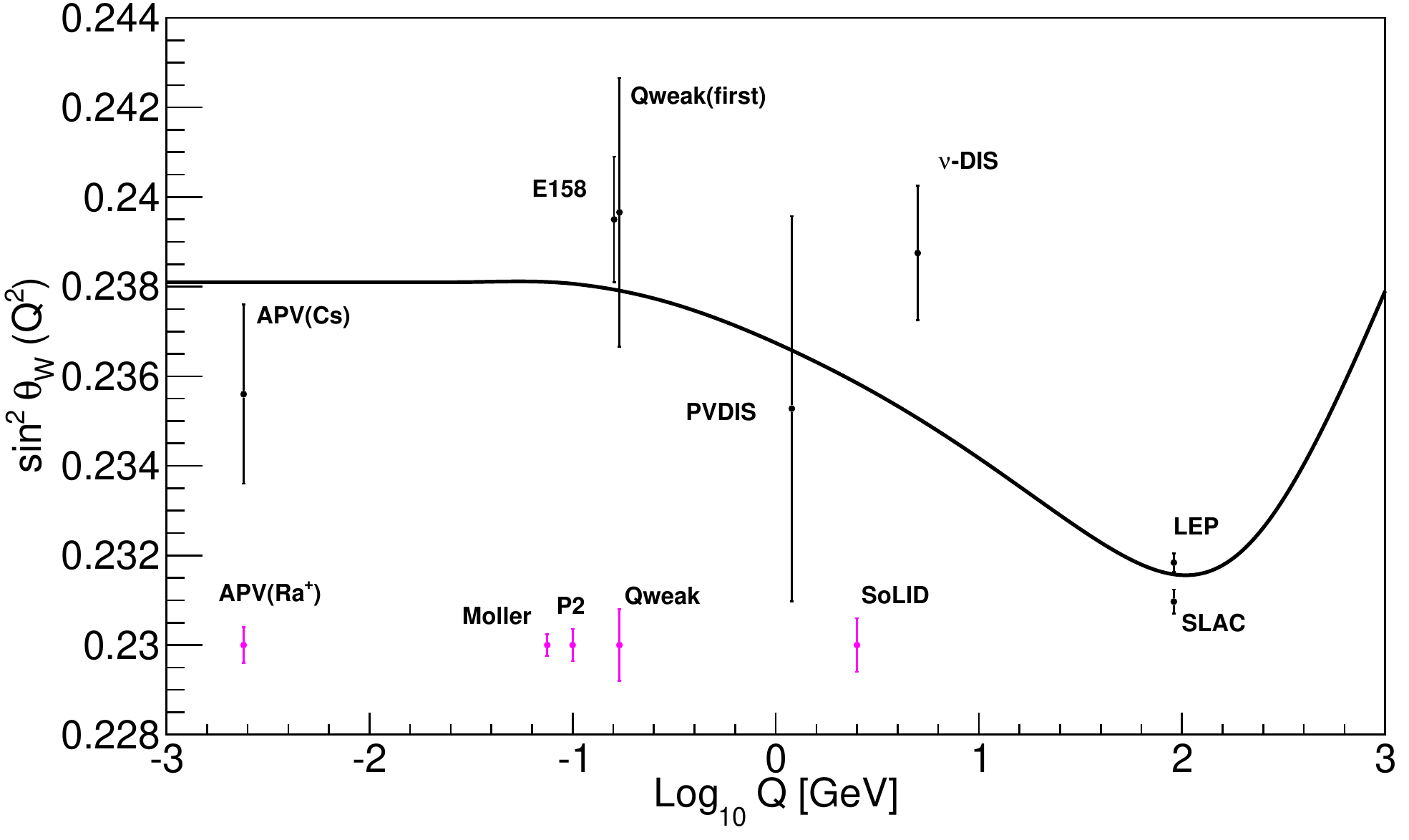}
\caption{(Color online) The $\sin^2 \theta_W$ projection from SoLID project along with other existing 
or proposed measurements \cite{PhysRevD.92.055005}. }
\label{fig:sin_world}
\end{figure}

In the context of the SM, the PVDIS in a $Q^2 \ll M_Z^2$ region with one-photon or one-$Z^0$ exchange between
the electron and the target can be expressed as \cite{xiaochao_nature}
\begin{equation}
A_{PV}=\frac{G_F Q^2}{4 \sqrt{2} \pi \alpha}[a_1(x) + a_3(x) \frac{1-(1-y)^2}{1+(1-y)^2}],
\label{eqn:APV}
\end{equation}
where $G_F$ is the Fermi constant, $\alpha$ is the electromagnetic fine-structure constant,
$Q^2$ is the squared momentum transfer to the electron,
$x$ is the Bjorken variable, 
$y$ is the fractional energy loss of the incident electron. 
The $a_{1,3}$ terms are
\begin{equation}
a_1(x)=2 g_A^e \frac{F_1^{\gamma Z}}{F_1^{\gamma}},
\end{equation}
\begin{equation}
a_3(x)=g_V^e \frac{F_3^{\gamma Z}}{F_1^{\gamma}}.
\end{equation}
The $F_{1,3}^{\gamma Z}$ functions are $\gamma-Z$ interference structure functions. 
In the parton model at the leading order, they can be written as:
\begin{equation}
F_1^{\gamma Z} = \sum_f e_{q_f} (g_V)_{q_f} (q_f + \bar{q}_f),
\end{equation}
\begin{equation}
F_3^{\gamma Z} = 2\sum_f e_{q_f} (g_A)_{q_f} (q_f-\bar{q}_f).
\end{equation}
The vector couplings $g_V$ of quarks and electrons are a function of $\sin^2 \theta_W$. For an iso-scalar target, such as a
deuteron in the valence region, which carries the same amount of $u$ and $d$ quarks, the contributions from PDFs cancel
in ratio in $a_{1,3}$ terms, hence the $A_{PV}$ is sensitive to $\sin^2 \theta_W$ directly: $A_{PV} \approx \frac{20}{3} \sin^2 \theta_W - 1$.
Figure \ref{fig:sin_world} shows the $\sin^2 \theta_W$ projection from SoLID 
along with other existing and proposed measurements.

\begin{figure}%[ht]
\includegraphics[width=90mm]{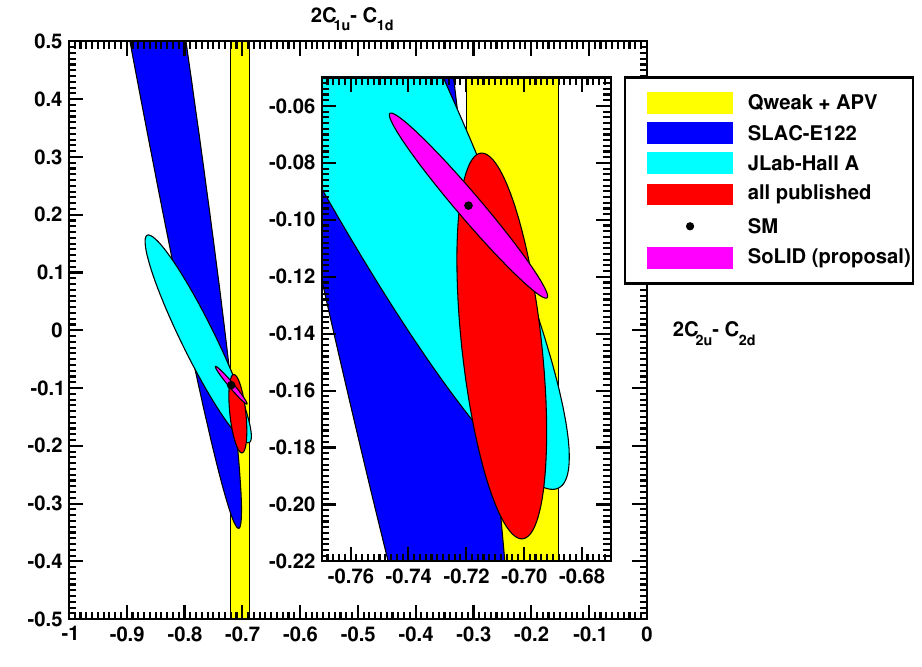}
\caption{(Color online) The phase-space of the linear combinations of axial-vector and vector-axial electron-quark effective
coupling constants for existing measurements \cite{xiaochao_nature} and a projection including measurements from SoLID project.} 
\label{fig:c1c2_solid}
\end{figure}

In the context of new physics searches, PVDIS can not be described only by the one-boson exchange. The effective electron-quark couplings
in terms of individual $g_A$ and $g_V$ are not valid anymore. Instead, the effective weak coupling constants $C_{1q,2q}$ are used.
In the leading order of one-boson exchange, they correspond to \cite{Wang:2014guo}:
\begin{eqnarray}
C_{1u} &=& 2 g_A^e g_V^u, ~~ C_{2u} = 2 g_V^e g_A^u, \\
C_{1d} &=& 2 g_A^e g_V^d, ~~ C_{2d} = 2 g_V^e g_A^d,
\end{eqnarray}
where $g_A$ and $g_V$ are the axial and vector couplings of electrons and up/down quarks.
If one neglects sea quarks in the valence region, then
\begin{equation}
a_1 = \frac{6}{5}(2C_{1u} - C_{1d}), a_3 = \frac{6}{5}(2C_{2u}-C_{2d}).
\end{equation}
At large $y$, $A_{PV}$ is sensitive to the $C_{2q}$, the coupling that can't be studied in low energy reactions due to large and uncertain
radiative corrections. Figure \ref{fig:c1c2_solid} shows existing and expected results on linear combinations of electron-quark weak coupling
constants for existing measurements and a projection after including measurements from SoLID proposal.

\begin{figure}%[ht]
\includegraphics[width=80mm]{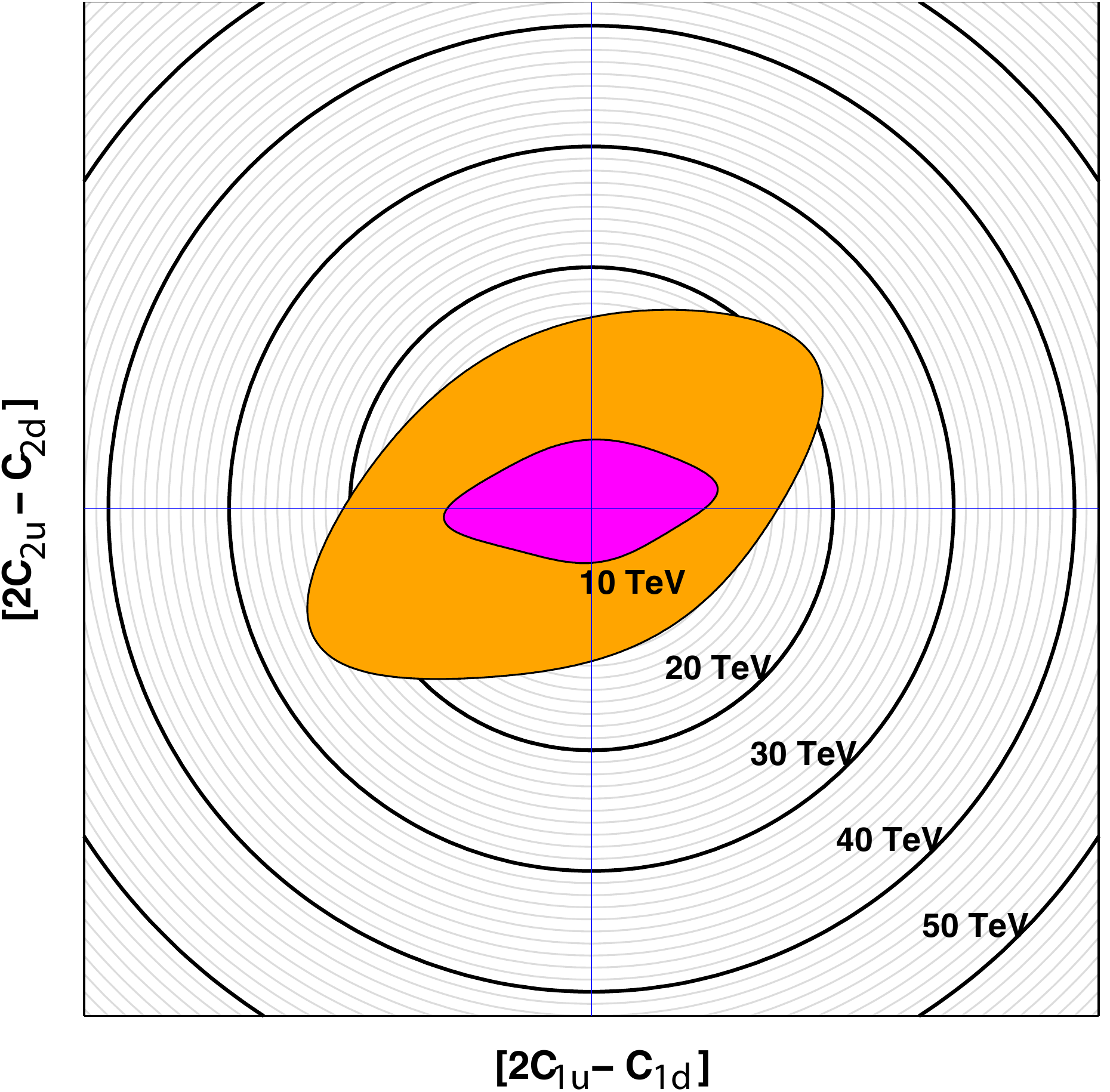}
\caption{(Color online) Mass-exclusion plot of the mass scales of new contact interactions assuming a physics coupling strength of $g^2=4 \pi$. The
pink (inner) region illustrates the reach by combining the 6 GeV PVDIS experiment at JLab and other precision experiments \cite{xiaochao_nature}, the orange (outer)
region shows the new reach assuming final precision from Qweak \cite{Androic:2013rhu} and SoLID PVDIS.} 
\label{fig:c1c2_mass}
\end{figure}

By measuring $C_{1q,2q}$, one can set constraints on new contact interactions, such as a possible lepto-phobic $Z$ boson.
To quantify and compare the physics reach of various experiments, one can quote mass limits within composite models \cite{Eichten:1983hw}, 
where the couplings are on the order of $4\pi/ \Lambda^2$ with $\Lambda$ the compositeness mass scale. The limit can be
extended to $\sim 20$ TeV level with the proposed precision of the SoLID proposal E12-10-007 and other existing measurements, as shown in Figure \ref{fig:c1c2_mass}. 

Other interesting topics in the high precision measurements of $A_{PV}$ in electron-deuteron scattering are the charge symmetry violation (CSV) 
and higher twist effects from quark-quark correlations. The strategy for the experiment is to have precision measurements over a broad kinematic range 
in both $x$ and $Q^2$. The data is fitted with the form
\begin{equation}
A_{\mathrm{measure}}=A_{SM}[ 1 + \frac{\beta_{HT}}{(1-x)^3} Q^2 + \beta_{CSV} x^2],
\end{equation}
where $\beta_{HT}$ is the asymmetry due to higher twist effects with a kinematic dependence of $\frac{1}{(1-x)^3 Q^2}$, $\beta_{CSV}$ is the asymmetry
due to CSV with a kinematic dependence of $x^2$. The Table \ref{table:kine} shows the kinematic sensitivities for different physics topics discussed above.

\begin{table}
\begin{center}
\begin{tabular}{|c|c|c|c|}\hline
	          &      $x$              &    $y$    &  $Q^2$           \\      \hline
New Physics       &     no                &     yes   &   small      \\ \hline
CSV               &    yes                &     small &   small       \\ \hline
High Twist        &   large?              &     no    &   large      \\ \hline
\end{tabular}
\caption{Kinematic dependence for different physics topics.}
\label{table:kine}
\end{center}
\end{table}

\subsection{PVDIS with longitudinally polarized electrons and a unpolarized proton target}
By measuring $A_{PV}$ from a unpolarized proton target, one can have direct access to the PDF $d/u$ ratio
free of nuclear effects. In the SM context at leading order and leading twist, $A_{PV}$ in Eq.
(\ref{eqn:APV}) is a function of $d/u$ in the valence quark region:
\begin{equation}
A_{PV} \approx - \frac{G_F Q^2}{2\sqrt{2} \pi \alpha} \frac{1+d/u}{4+d/u},
\end{equation}
where it has been assumed that $\sin^2 \theta_W = \frac{1}{4}$. The traditional way of determining $d/u$ relies on comparing the
inclusive DIS cross section on a proton target to that of a deuteron target. 
The disadvantages, compared to the asymmetry measurement, are that the cross section measurement 
is hard to achieve with high precisions and nuclear corrections in the deuteron target in the large $x$ region 
lead to large uncertainties.

\begin{figure}%[ht]
\includegraphics[width=90mm]{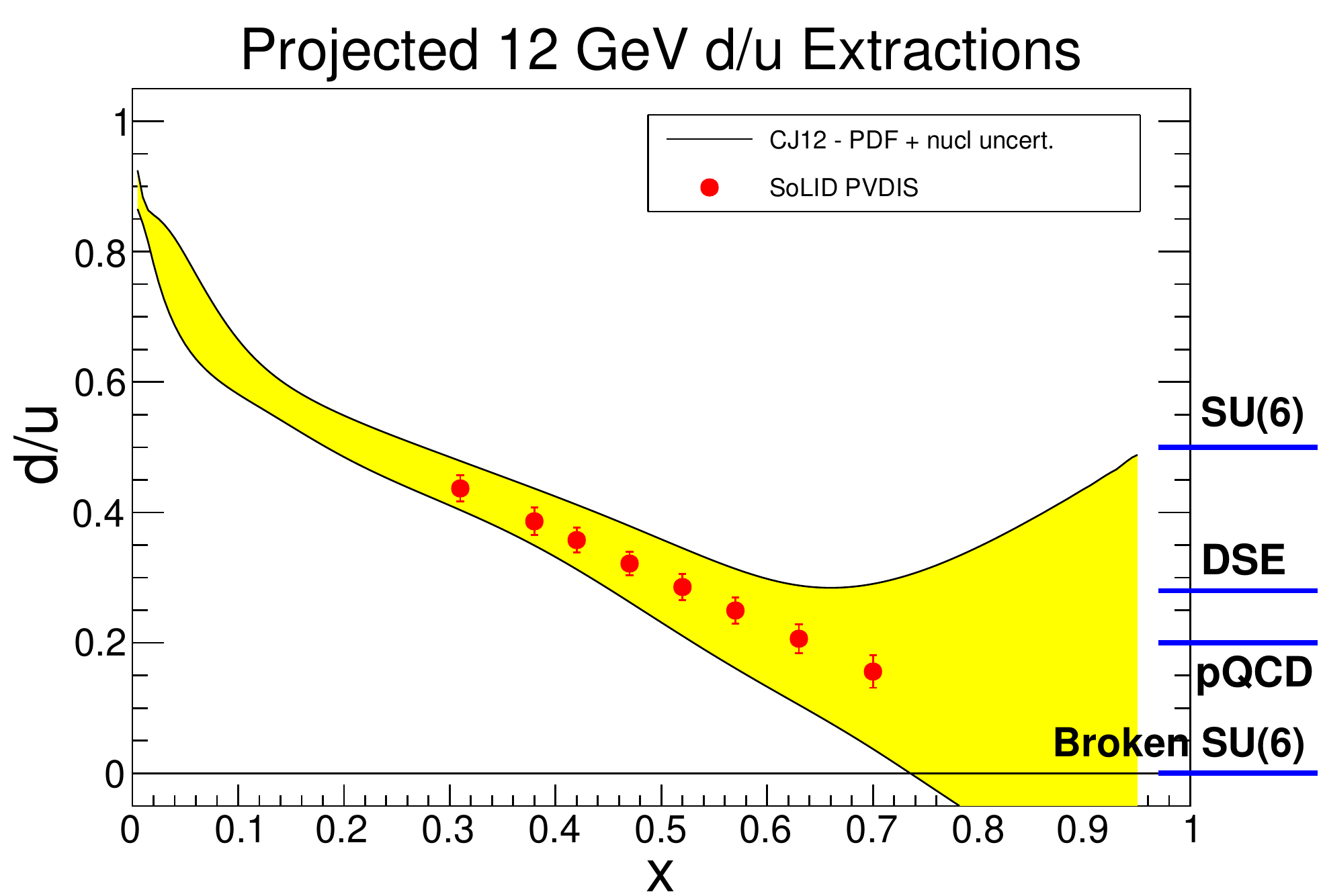}
\caption{(Color online) The projections of $d/u$ ratio from SoLID proposal, also shown are the calculations
based on PDFs from the CJ12 collaboration \cite{Accardi:2016qay}.}
\label{fig:d_over_u}
\end{figure}

The projections on $d/u$ from SoLID proposal E12-10-007 is shown in Figure \ref{fig:d_over_u} along with calculations
using PDFs from the CJ12 collaboration. The data from SoLID will be complementary to other proposed experiments at JLab including
the one using $^3$H and $^3$He nuclei to minimize nuclear effects during $d/u$ extraction \cite{Arrington:2011qt}, 
and the BoNuS experiment \cite{Baillie:2011za} at JLab Hall B.

\subsection{PVDIS with unpolarized electrons and a longitudinally polarized $^3$He target}
Another attractive physics topic with PVDIS is to use a longitudinally polarized nuclear target and unpolarized electrons. In the DIS region, 
the single-target parity violation asymmetry is determined by the polarized electroweak interference structure function
$g_{1,5}^{\gamma Z}$ of the nucleon \cite{Anselmino:1994gn}:
\begin{equation}
A_{L} = \frac{\sigma^{(+)} - \sigma^{(-)}}{\sigma^{(+)} + \sigma^{(-)}} = \frac{G_F Q^2}{2\sqrt{2} \pi \alpha}[g_V^e \frac{g_5^{\gamma Z}}{F_1^{\gamma}} + g_A^e \frac{2y-y^2}{y^2-2y+2} \frac{g_1^{\gamma Z}}{F_1^{\gamma}}],
\end{equation}
where 
\begin{eqnarray}
g_1^{\gamma Z} &=& \sum_f e_{q_f} (g_V)_{q_f}(\Delta q_f +\Delta\bar{q}_f), \\
g_5^{\gamma Z} &=& \sum_f e_{q_f} (g_A)_{q_f}(\Delta q_f - \Delta\bar{q}_f).
\end{eqnarray}
If one assumes $\sin^2 \theta_W = \frac{1}{4}$, the $g_1^{\gamma Z}$ function is approximately proportional
to $\Delta \Sigma \equiv \sum_{f} (\Delta q_f + \Delta \bar{q}_f)$, with $\Delta q_f$ the polarized parton
distribution functions. The $g_5^{\gamma Z}$ function is sensitive to the valence quark polarization
$\Delta q_V \equiv \Delta q - \Delta \bar{q}$, including $\Delta s- \Delta \bar{s}$ that can't
be measured from existing experimental techniques.
With the input of the weak mixing angle in the context of the SM, these brand new and yet unmeasured polarized electroweak interference structure functions
can be extracted, providing independent information on the spin structure of the nucleon 
and SU(3) flavor symmetry test in addition to the $g_1^{\gamma}$ structure function.

\begin{figure}%[ht]
\includegraphics[width=90mm]{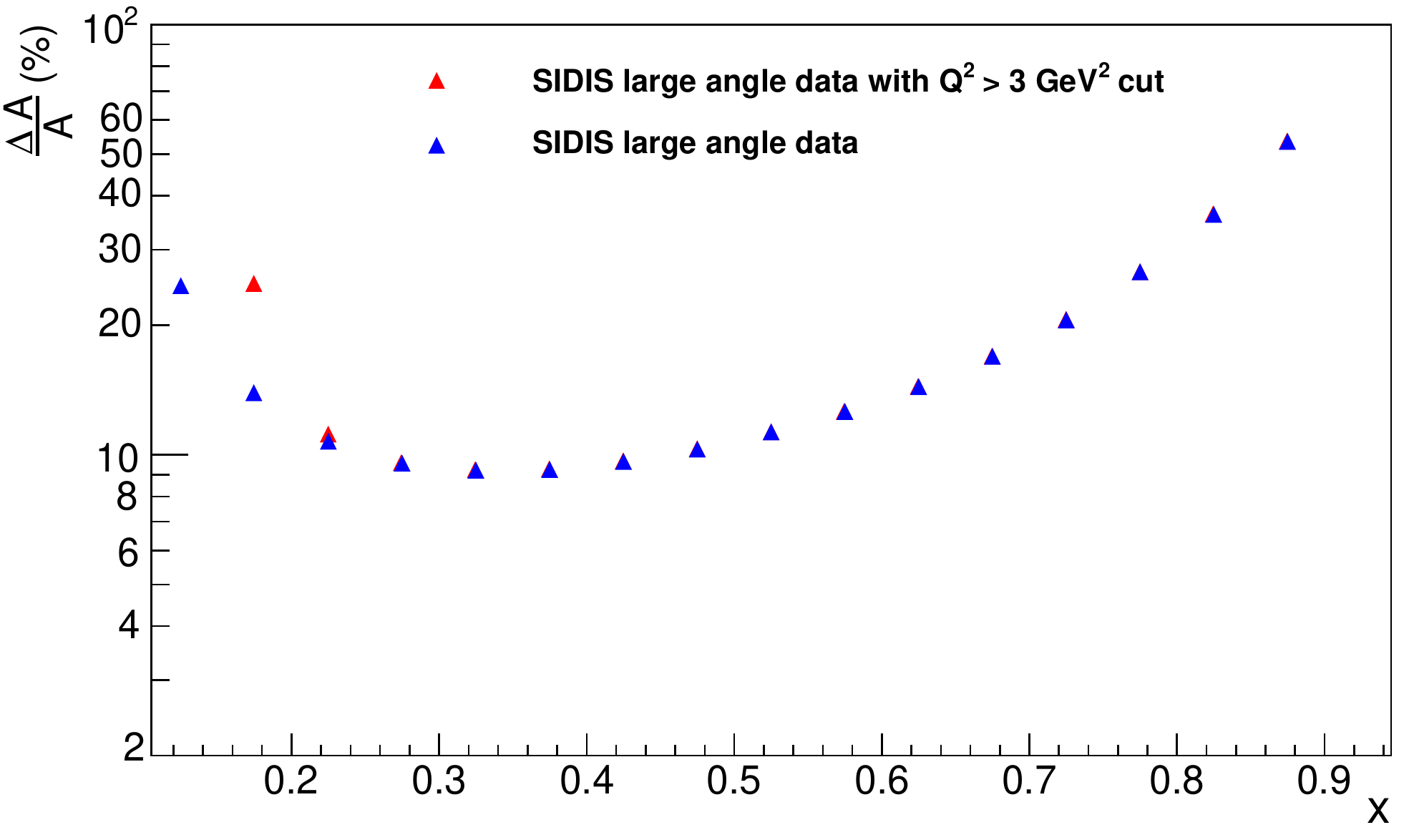}
\caption{(Color online) The expected uncertainty on the $^3$He single-target PVDIS asymmetry. The $Q^2$ cut only affects the last two data points in 
$x<0.25$ region as illustrated by the red points. }
\label{fig:polAPV}
\end{figure}

A letter-of-intent was submitted to JLab PAC44 (LOI-12-16-007) to carry out the first such measurement with 
a longitudinally polarized $^3$He target at SoLID. The predicted relative uncertainties 
on $A_{PV}$ measurements as a function of $x$ are shown in Figure \ref{fig:polAPV}.

\subsection{PVDIS with longitudinally polarized electrons and a unpolarized $^{48}$Ca target}
PVDIS on a heavy nuclear target will provide a direct measurement of flavor-dependent nuclear medium modification effects on quarks. 
The $a_1$ term in Eq. (\ref{eqn:APV}) for a nuclear target with atomic number $A$ can be written as
\begin{equation}
a_1 \approx \frac{9}{5} - 4 \sin^2 \theta_W - \frac{12}{25} \frac{u_A^+- d_A^+}{u_A^+ + d_A^+},
\end{equation}
with the convention that $q_{A}^{\pm} = q_A(x) \pm \bar{q}_A(x)$. Therefore, the measurement is directly sensitive to 
differences in the quark flavors within a nucleus, which would represent new and important information on our
understanding of the EMC effect. The flavor dependence can also be used to examine different models for the 
scaling of the EMC effect \cite{Arrington:2015wja}, for instance the idea that the EMC effect scales with the large
virtuality of the struck nucleon or those which scale with the nucleon's local density \cite{Arrington:2012ax}.

\begin{figure}%[ht]
\includegraphics[width=90mm]{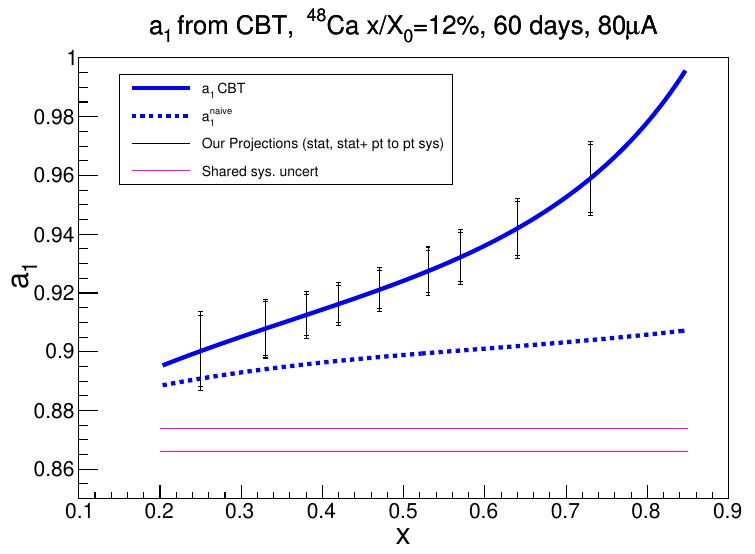}
\caption{(Color online) The projections for the $a_1$ function using a $^{48}$Ca target. The calculations using the CBT model and 
PDFs are also shown.}
\label{fig:APVCa}
\end{figure}

A measurement requesting 60 days of beam time with 80$\mu$A beam current on a $^{48}$Ca target was proposed (PR12-16-006). The predictions for 
$a_1$ as a function of $x$ are shown in Figure \ref{fig:APVCa}. The prediction from the CBT model \cite{Cloet:2009qs} is also shown in the plot. 
The CBT model has been very successful in reproducing the quark distributions for the EMC effect as well as the measured structure functions.
It is also able to explain part of the NuTeV anomaly.

\section{The Full SoLID Program}
In addition to the PVDIS program discussed above, there are other rich programs based on the SoLID spectrometer \cite{Chen:2014psa}. The detector subsystems
can be reconfigured to accommodate Semi-Inclusive Deep Inelastic Scattering (SIDIS) with polarized $^3$He and proton targets
to measure Transverse Momentum Dependent Parton Distributions (TMDs) in multi-dimensional kinematics with high precisions \cite{Jlab12GeV}. There is
also a $J/\psi$ physics program to study the threshold electroproduction of the $J/\psi$ on the nucleon, which provides a unique
opportunity to help understand the low energy struture of the nucleon. Recently, there are more ideas to carry out
Deeply Virtual Compton Scattering (DVCS) programs with SoLID. The unique feature of combining the large acceptance and high luminosity of SoLID
makes it critical to exploit the full potential of the JLab 12 GeV upgrade to perform precision studies of the nucleon structure and 
QCD dynamics.

\section{Summary}
The SoLID spectrometer will provide the opportunities to measure the weak mixing angle and effective
electron-quark weak couplings $C_{1q,2q}$, especially $C_{2q}$, to very high precisions. There is also potential to access
the CSV and high twist effects with quark-quark correlations.
With a proton target, the measurement of the $d/u$ ratio can be achieved in a relatively clean way, free of nuclear corrections. PVDIS
with a polarized target opens up a new window to access the new, yet unmeasured polarized $\gamma Z$ interference structure 
functions, and provides independent inputs on unique combinations of polarized PDFs to the world PDF fit. The SU(3)
flavor symmetry could also be tested. Furthermore, by using a $^{48}$Ca target, one can observe a possible isovector EMC effect
that can help understand the NuTeV anomaly. The SoLID spectrometer, which is designed to accomodate
a high luminosity with a large acceptance, also provides unique opportunities to various physics programs including SIDIS, $J/\psi$ and DVCS
experiments. More ideas are coming out in the near future within the collaboration. In summary, the SoLID project will greatly enhance
the physics output of the JLab 12 GeV upgrade in a number of exciting areas.

\section*{Acknowledgements}
The author is grateful to the inputs from Jian-Ping Chen, Krishna Kumar, Seamus Riordan, Paul Souder, Xiaochao Zheng for this article.
The material is based upon the work supported by the U.S. Department of Energy, Office of Science, Office of Nuclear Physics under
contract DE-AC05-06OR23177. It is also supported in part by U.S. Department of Energy, Office of Science under contract DE-FG02-84ER40146.

\nocite{*}
\bibliography{references}

\end{document}